\documentclass[twocolumn,showpacs,amsmath,amssymb,prl]{revtex4}
\usepackage{graphicx}
\usepackage{dcolumn}
\usepackage{bm}
\begin{document}
\title{Elastic Behavior of a Two-dimensional Crystal near Melting}
\author{H.H. von Gr\"unberg}
\affiliation{Karl-Franzens-Universit\"at, 8010 Graz, Austria}
\author{P. Keim, K. Zahn, G. Maret}
\affiliation{Universit\"at Konstanz, 78457 Konstanz, Germany}
\date{\today}
\begin{abstract}
  Using positional data from video-microscopy we determine the elastic
  moduli of two-dimensional colloidal crystals as a function of
  temperature. The moduli are extracted from the wave-vector-dependent
  normal mode spring constants in the limit $q\to 0$ and are compared
  to the renormalized Young's modulus of the KTHNY theory. An
  essential element of this theory is the universal prediction that
  Young's modulus must approach $16 \pi$ at the melting
  temperature. This is indeed observed in our experiment.
\end{abstract}
\maketitle

In the early 70th Kosterlitz and Thouless \cite{Kosterlitz73}
developed a theory of melting for two-dimensional systems. In their
model the phase transition from a system with quasi-long-range order
\cite{mermin} is mediated by the unbinding of topological defects like
vortices or dislocation pairs in the case of 2D crystals. They showed
that the phase with higher symmetry has short-range translational
order.  Halperin and Nelson \cite{Halperin78,nelson79} pointed out
that this phase still exhibits quasi-long-range orientational order
and proposed a second phase transition now mediated by the unbinding
of disclinations to an isotropic liquid. The intermediate phase is
called hexatic phase. This theory, being based also on the work of
Young \cite{Young79}, is known as KTHNY theory (Kosterlitz, Thouless,
Halperin, Nelson and Young); it describes the temperature-dependent
behavior of the elastic constants, the correlation lengths, the
specific heat and the structure factor (for a review see
\cite{Strandburg88, Glaser93}).  Experiments with electrons on helium
\cite{Grimes79,Morf79} and with 2D interfacial colloidal systems
\cite{Pieranski80,Murray87,Kusner94,Marcus97,zahn99} as well as
computer simulations \cite{Chen95,Bagchi96,Jaster98} have been
performed to test the essential elements of this theory. But research
has mainly focused on the behavior of the correlation functions (an
illustrative example is the work of Murray and van Winkle
\cite{Murray87}). Only a few works can be found that deal with the
elastic constants, especially the shear modulus,
\cite{Morf79,Fisher82,Morales94,Sengupta00} even though the
Lam$\acute{e}$ coefficients and their renormalization near melting
take a central place in the KTHNY theory.

A very strong prediction of the KTHNY theory has never been verified
experimentally. It states that the renormalized Young's Modulus
$K_R(T)$, being related just to the renormalized Lam$\acute{e}$
coefficients $\overline{\mu}_{R}$ and $\overline{\lambda}_{R}$, must
approach the value $16\pi$ at the melting temperature \cite{nelson79},
\begin{equation}
  \label{eq:0}
  K_{R}(T) = 4 \overline{\mu}_{R} (1 -
\overline{\mu}_{R}/(2 \overline{\mu}_{R} + \overline{\lambda}_{R}))
 \to 16 \pi \quad \mbox{if}\: T \to T_m^-\:,
\end{equation}
which is obviously an universal property of 2D systems at the melting
transition. This Letter presents experimental data for elastic moduli
of a two-dimensional colloidal model system, ranging from deep in the
crystalline phase via the hexatic to the fluid phase. These data,
indeed, confirm the theoretical prediction expressed by
eq.~(\ref{eq:0}).

The experimental setup is the same as already described in
\cite{keim04}. The system is known to be an almost perfect 2D system;
it has been successfully tested, and explored in great detail, in a
number of studies \cite{zahn99,zahn97,zahn00,keim04}.  Therefore we
only briefly summarize the essentials here: Spherical colloids
(diameter $d=4.5 \:\mu m$) are confined by gravity to a water/air
interface formed by a water drop suspended by surface tension in a top
sealed cylindrical hole of a glass plate. The flatness of the
interface can be controlled within $\pm$ half a micron. The field of
view was $835 \times 620 \: \mu m^{2}$ containing typically up to
$3\cdot10^{3}$ particles (the whole system has a size of $50~mm^{2}$
and contains about $3\cdot10^{5}$ particles). The particles are
super-paramagnetic, so a magnetic field $\vec{B}$ applied
perpendicular to the air/water interface induces in each particle a
magnetic moment $\vec{M}= \chi \vec{B}$ which leads to a repulsive
dipole-dipole pair-interaction energy of $\beta v(r) = \Gamma/
(\sqrt{\pi\rho}r)^{3}$ with the dimensionless interaction strength
given by $\Gamma = \beta (\mu_{0}/4 \pi) (\chi B)^{2} (\pi
\rho)^{3/2}$ ($\beta = 1/kT$ inverse temperature, $\chi$
susceptibility, $\rho$ area density). The interaction can be
externally controlled by means of the magnetic field $B$. $\Gamma$ was
determined as in Ref.~\cite{zahn99} and is the only parameter
controlling the phase-behavior of the system. It may be considered as
an inverse reduced temperature, $T = 1/\Gamma$. For $\Gamma >
\Gamma_m=60$ the sample is a hexagonal crystal \cite{zahn99,zahn00}.
Coordinates of all particles at equal time steps and for different
'temperatures', i.e. $\Gamma$'s, were recorded using digital
video-microscopy and evaluated with an image-processing software.  We
measured over 2-3 hours and recorded trajectories of about 2000
particles in up to 3600 configurations, for a large number of
different $\Gamma$'s ranging between $\Gamma = 49$, deep in the fluid
phase, to $\Gamma =175$ in the solid phase. These trajectories were
then further processed to compute the elastic constants of the
colloidal crystal as a function of the inverse temperature $\Gamma$.

Our data analysis is based on the classical paper of Nelson and
Halperin (NH) on dislocation-mediated melting in 2D systems
\cite{nelson79}. Their considerations start from the reduced elastic
Hamiltonian,
\begin{equation}
  \label{eq:1}
  \beta {\cal H}_{E} = \frac{1}{2} \int \frac{d^{2}r}{a^{2}}
\big[2 \overline{\mu} u_{ij}^{2} +\overline{\lambda} u_{kk}^{2} \big]
\end{equation}
where $a$ is the lattice constant of a triangular lattice (next
neighbor distance), while $\overline{\mu} = \mu a^{2}/kT$ and
$\overline{\lambda}= \lambda a^{2}/kT$ denote the dimensionless
Lam$\acute{e}$ coefficients ($\beta = 1/kT$).  $u_{ij}(\vec{r}) =
(\partial_{r_{j}} u_{i}(\vec{r}) +
\partial_{r_{i}}u_{j}(\vec{r}))/2$ is the usual strain tensor related to
the displacement field $\vec{u}(\vec{r})$.  At temperatures $T$ near
the melting temperature $T_m$ the field $u_{ij}(\vec{r})$ contains
singular parts $u^{sing}_{ij}(\vec{r})$ due to dislocations; it can be
decomposed into $u_{ij}(\vec{r})= u^{sing}_{ij}(\vec{r}) +
\phi_{ij}(\vec{r})$ with $\phi_{ij}(\vec{r})= (\partial_{r_{j}}
\phi_{i}(\vec{r}) + \partial_{r_{i}}\phi_{j}(\vec{r}))/2$ being a
smoothly varying function ($\vec{\phi}(\vec{r})$ is the regular part
of the displacement field $\vec{u}(\vec{r})$). When this decomposition
is inserted in eq.~(\ref{eq:1}), the Hamiltonian decomposes into two
parts, $\beta {\cal H}_{E} = 1/2 \int d^{2}r/a^{2} (2 \overline{\mu}
\phi_{ij}^{2} +\overline{\lambda} \phi_{kk}^{2}) + \beta {\cal H}_{D}$
with ${\cal H}_{D}$ representing the extra elastic energy that is due
to the dislocations. NH were able to derive a set of differential
equations for renormalized Lam$\acute{e}$ coefficients,
$\overline{\mu}_{R}$ $\overline{\lambda}_{R}$, by means of which
${\cal H}_E$ can again be written as in eq.~(\ref{eq:1}),
\begin{equation}
  \label{eq:3}
  \beta {\cal H}_{E} = \frac{1}{2} \int \frac{d^{2}r}{a^{2}}
\big[2 \overline{\mu}_{R} u_{ij}^{2} +\overline{\lambda}_{R} u_{kk}^{2} \big]
\:.
\end{equation}
Because the effect of the dislocations are entirely absorbed into the
elastic constants, the strain tensor in eq.~(\ref{eq:3}) can now be
assumed to be again regular everywhere and for all $T<T_m$.

Our experiment measures the trajectories $\vec{r}_{i}(t)$ of $N$
particles of a colloidal crystal over a finite time window of width
$t_{exp}$. Associating the average $\langle \vec{r}_{i}
\rangle_{t_{exp}}$ with a lattice site $\vec{R}_{i}$, we can, for each
particle, compute displacement vectors $\vec{u}(\vec{R}_{i}) =
\vec{r}_{i} - \vec{R}_{i}$ . The Fourier transforms of these
displacement vectors, $\vec{u}(\vec{q}) = N^{-1/2} \sum_{\vec{R}} e^{i
\vec{q}\vec{R}} \vec{u}(\vec{R})$, are now used for the numerical
computation of renormalized elastic constants. This has been done in
the following way. Starting from eq.~(3.29) of the NH paper,
\begin{equation}
  \label{eq:4}
\lim_{\vec{q}\to0} q^2
\langle u^\ast_i(\vec{q})  u_j(\vec{q}) \rangle
=
\frac{kT}{v_0}
\Big[
\frac{1}{\mu_R} \delta_{ij} -
\frac{\mu_R + \lambda_R}{\mu_R(2\mu_R + \lambda_R)}
\frac{q_i q_j}{q^2}
\Big]
\end{equation}
we find, after decomposing the displacement field $\vec{u}(\vec{q})$
into parts $\vec{u}_{||}(\vec{q})$ and $\vec{u}_{\bot}(\vec{q})$,
parallel and perpendicular to $\vec{q}$, that
\begin{eqnarray}
  \label{eq:5} \frac{v_0(2\mu_R + \lambda_R)}{kT}
& = & \lim_{\vec{q}\to0} \zeta_{||}(\vec{q}), \:\:
\zeta_{||}(\vec{q}):= \big( q^2
\langle |u_{||}(\vec{q})|^2 \rangle \big)^{-1}\\
  \label{eq:6} \frac{v_0 \mu_R}{kT} & = &
\lim_{\vec{q}\to0}  \zeta_{\bot}(\vec{q}),\:\:
\zeta_{\bot}(\vec{q}):= \big( q^2
\langle |u_{\bot}(\vec{q})|^2\rangle \big)^{-1}
\end{eqnarray}
where $v_{0}=\sqrt{3}a^2/2$ is the area per colloid in a triangular
lattice.

These two equations are the central equations for our data evaluation
scheme, and are therefore more carefully discussed. Deep in
the solid phase $\langle \vec{r} \rangle_{\Delta t}$ converges with
increasing measurement time $\Delta t$ to lattice sites, and $\langle
|\vec{u}(\vec{R})|^2 \rangle$ and $(q^2 \langle |\vec{u}(\vec{q})|^2
\rangle)_{q\to 0}$ remain finite, and lead thus to two non-zero
elastic moduli: the shear modulus $\mu$ and the bulk modulus
$B=\lambda + \mu$. By contrast, in the fluid phase $\langle \vec{r}
\rangle_{\Delta t}$ will neither converge to, nor correlate with, any
sort of lattice site and, in addition, as the mean square displacement
is unbound, $(q^2 \langle|\vec{u}(\vec{q})|^2 \rangle)_{q\to 0} \to
\infty$, so, both moduli will vanish (while $B$
survives \cite{foot37HN}).

Though reasonable at first glance, this interpretation of
eqs.~(\ref{eq:5},\ref{eq:6}) is in fact an oversimplification and
ignores the limited applicability of both equations. This can best be
seen by re-deriving them, first by Fourier-transforming
eq.~(\ref{eq:3}), by then inserting the decomposition
$\vec{u}=\vec{u}_{||}+ \vec{u}_{\bot}$ and by applying finally the
equipartition theorem. The intimate relationship between
eq.~(\ref{eq:3}) and eqs.~(\ref{eq:5},\ref{eq:6}) let us realize that
the $\vec{u}(\vec{q})$ in eqs.~(\ref{eq:5},\ref{eq:6}) refer to a
coarse-grained and thus regular displacement field, just as in
eq.~(\ref{eq:3}). In other words, with eqs.~(\ref{eq:5},\ref{eq:6})
the softening of the elastic constants for $T \to T_m^-$ is inferred
indirectly, namely from the change of behavior of the regular parts of
a coarse-grained displacement field.

We here identify this coarse-grained displacement field with the
$\vec{u}(\vec{r}(t))=\langle \vec{r} \rangle_{t_{exp}} - \vec{r}(t)$
evaluated from our experimental data. In doing so, we have to be aware
of the following two points. (i) In an imperfect crystal, especially
in the presence of dislocations, $\langle \vec{r} \rangle_{t_{exp}\to
  \infty}$ does not always converge to lattice sites.  As we are
interested in the limit $q\to 0$, this is unproblematic, as long as
these extra sites do not move. Looking at our measured trajectories,
we have observed only thermally activated dislocation pairs, but no
static, isolated dislocations traveling through the crystal. (ii) The
displacement field computed from the experimental data has (and must
have) parts stemming from dislocations. The resulting error should be
small below $T_m$ when the number of dislocations is still relatively
low (even at $T=T_{m}$, the probability that a particle belongs to a
dislocation is only 1 \% !).  However, the error should become
appreciable at $T>T_m$ where eqs.~(\ref{eq:5},\ref{eq:6}) can not be
expected to strictly hold any longer.

\begin{figure}
\includegraphics[width=0.48 \textwidth]{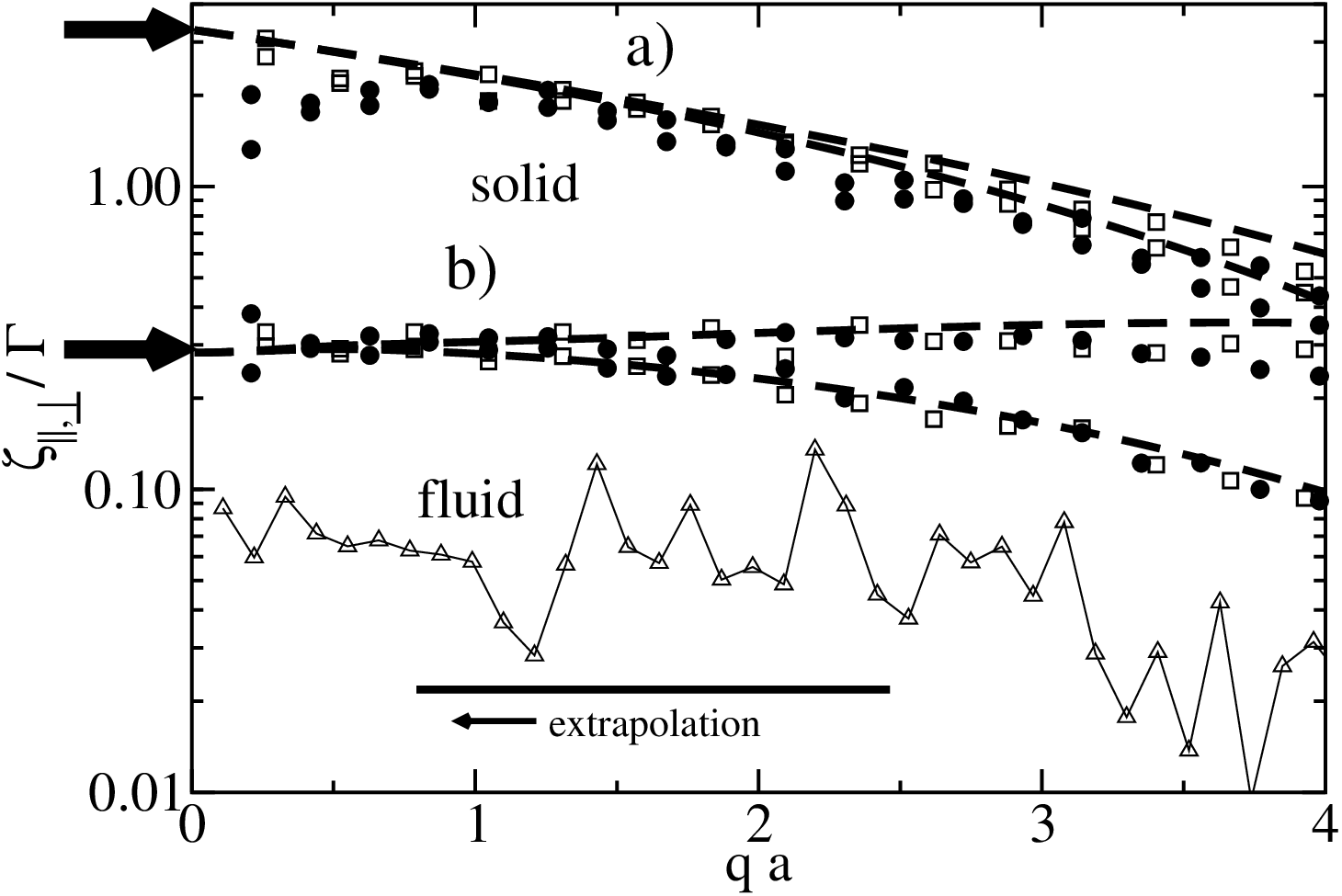}
\caption{\label{fig1}
  $\zeta_{||}(\vec{q})/\Gamma$ in a) and
  $\zeta_{\bot}(\vec{q})/\Gamma$ in b), as defined in
  eqs.~(\ref{eq:5},\ref{eq:6}).  Each quantity is plotted for two
  different directions in $q$-space ($\Gamma \to M$ and $\Gamma \to K$
  in the first Brillouin zone), and for three different values of the
  interaction-strength parameter $\Gamma$: $\Gamma = 52$ in the fluid
  phase (open triangles), $\Gamma =75$ (open squares), and
  $\Gamma=125$ (filled circles) in the crystalline phase. Thick solid
  arrows for a $T=0$ prediction of the elastic moduli, dashed solid
  lines for the predictions of harmonic crystal theory. For $\Gamma =
  52$ just one band is shown.}
\end{figure}
For the pair-potential $\sim \Gamma/r^{3}$ the elastic constants can
be calculated in the limit $\Gamma \to \infty$ ($T=0$) using simple
thermodynamical relations involving essentially lattice sums of the
pair-potential.  One finds $\overline{\lambda} + \overline{\mu} = 3.46
\:\Gamma$ and $\overline{\mu} = 0.346 \:\Gamma$ \cite{wille}. For
convenience, we divide in the following all moduli by $\Gamma$.
Fig.~(\ref{fig1}) shows $(\lambda+2 \mu)v_{0}/\Gamma k T$ and $\mu
v_{0}/\Gamma k T$, obtained from this $T=0$ calculation, as thick
solid arrows, and compares it to the expressions
$\zeta_{||}(\vec{q})/\Gamma$ and $\zeta_{\bot}(\vec{q})/\Gamma$ from
eqs.~(\ref{eq:5},\ref{eq:6}), as obtained from the measured
trajectories for three different values of $\Gamma$. Let us first
focus on the measurement for $\Gamma=75,125$ where the system is deep
enough in the crystalline phase for the assumption $T=0$ to be valid.
$\zeta_{||}(\vec{q})/\Gamma$ and $\zeta_{\bot}(\vec{q})/\Gamma$, indeed,
tend to the predicted elastic constants in the limit $q \to 0$, in
agreement with eqs.~(\ref{eq:5},\ref{eq:6}).  For either of the two
plotted quantities, we consider two different high-symmetry directions
in $q$ space, which are $\vec{q}_1=q\vec{b}_1$ and
$\vec{q}_2=q(\vec{b}_1 + \vec{b}_2)/\sqrt{2}$ where $\vec{b}_1=(1,0)a$
and $\vec{b}_2=(1,\sqrt{3})a/2$ are basis vectors of the (hexagonal)
reciprocal lattice. At wavelengths larger than the lattice constant
($qa<1$), the results for both bands are identical, thus indicating an
essentially isotropic $\vec{u}(\vec{q})$ at small $q$.
$\zeta_{||}(\vec{q})$ and $\zeta_{\bot}(\vec{q})$ can also
be associated with the $\vec{q}$-dependent normal-mode spring
constants (elastic dispersion curves) of the discrete crystal, having
longitudinal $\lambda_{long}(\vec{q})$ and transversal
$\lambda_{trans}(\vec{q})$ branches. This can be (and have been
\cite{keim04}) compared to the band-structure predicted by harmonic
lattice theory (thick dashed lines in Fig.~(\ref{fig1}), for details
see \cite{keim04}). In other words, what we do here is to derive
elastic constants from the $q\to 0$ behavior of the elastic
dispersions curves ($\lim_{q\to0} \lambda_{long}(\vec{q}) = (\lambda
+2 \mu) q^2 v_0$, $\lim_{q\to0} \lambda_{trans}(\vec{q}) = \mu q^2
v_0$).

While for all our measurements above $\Gamma_m = 60$ the resulting
bands lie on top of the dashed thick lines in Fig.~(\ref{fig1}), one
finds a systematic shift to smaller values for $\Gamma < \Gamma_m$.
Fig.~(\ref{fig1}) shows, as an example, one out of the four bands
(belonging to $\zeta_{\bot}(\vec{q})/\Gamma$) of the measurement in
the fluid phase ($\Gamma = 52$). It lies an order of magnitude below
the crystalline bands.

\begin{figure}
\includegraphics[width=0.48 \textwidth]{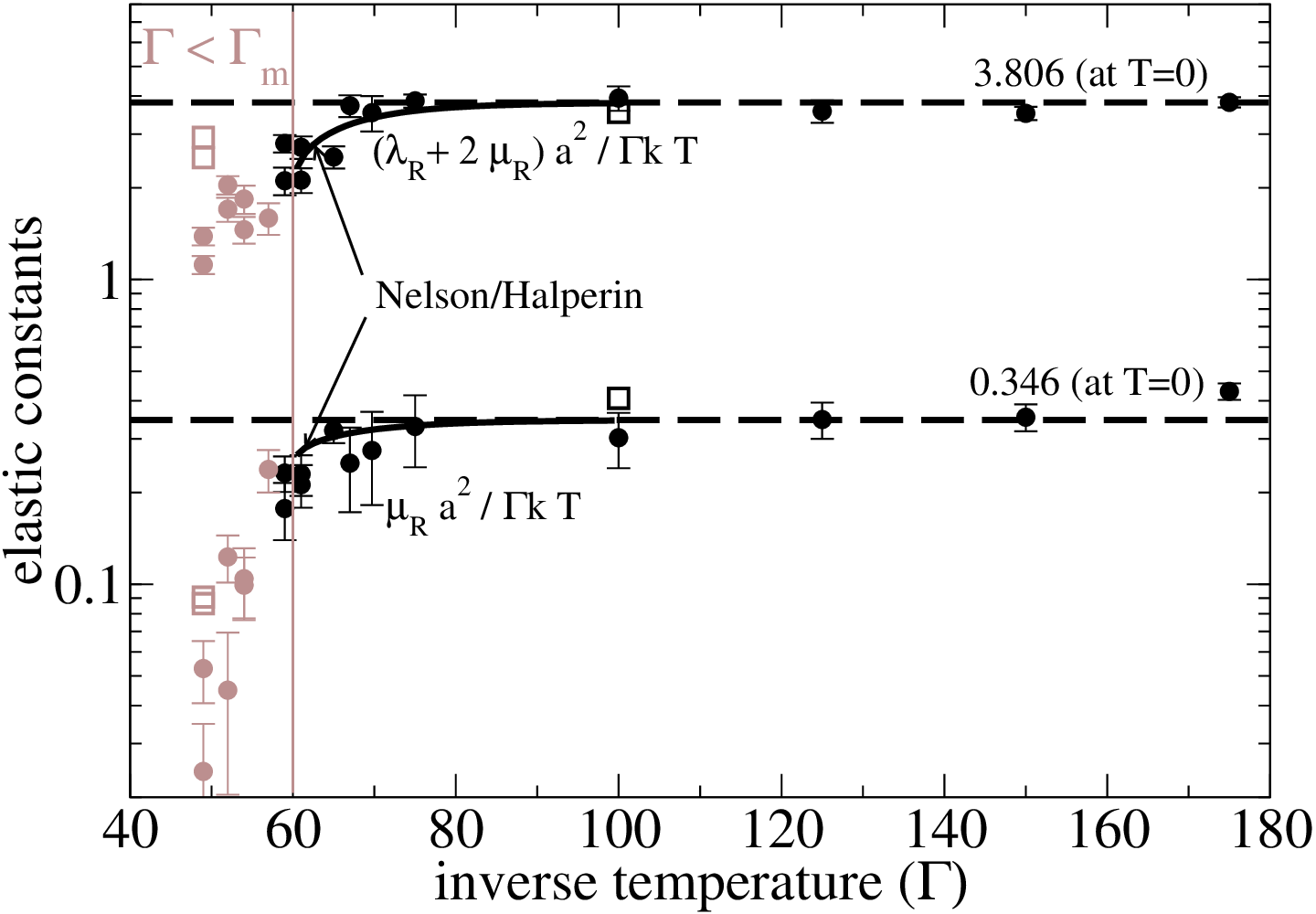}
\caption{\label{fig2}
  Elastic moduli of a 2D colloidal crystal as a function of the
  inverse temperature, obtained from extrapolating the bands in
  Fig.~(\ref{fig1}) down to $q=0$. The melting temperature is at
  $\Gamma_m = 60$.
  Thick dashed lines for a $T=0$ prediction of the elastic moduli,
  $\overline{\lambda} + 2 \overline{\mu} = 3.806 \Gamma$ and
  $\overline{\mu} = 0.346 \Gamma$ \cite{wille}; thick solid line for
  the theoretical elastic constants, renormalized as described in
  \cite{nelson79}.
}
\end{figure}

In order to infer functions $\mu_R(\Gamma)$ and $\lambda_R(\Gamma)$
from these bands, we need to take the limit $q \to 0$. Since at low
$q$ we have to expect finite size effects, and at high $q$, near the
edges of the first Brillouin zone, effects resulting from the band
dispersion of the discrete lattice, we choose an intermediate $q$
regime ($0.8<q a <2.5$), indicated by the thick solid bar in
Fig.~(\ref{fig1}), to extrapolate the bands to $q=0$, applying a
linear regression scheme.  The extrapolation procedure was optimized
at high $\Gamma$ for which we have a precise idea what constants we
should find.  For each modulus, extrapolation of the two bands
depicted in Fig.~(\ref{fig1}) were checked: while for
$\zeta_{||}/\Gamma$, one finds for both bands the same modulus, the
upper band of the two for $\zeta_{\bot} /\Gamma$ gave much better
result and was henceforth taken. Also, different extrapolation schemes
have been checked, but linear extrapolation turned out to produce a
tolerably small error, much smaller than the main error of our
measurement, estimated here from the standard error in the linear
regression scheme.

Fig.~(\ref{fig2}) shows the resulting moduli, for all values of
$\Gamma$ studied. Black symbols refer to systems in the crystalline
state ($\Gamma>\Gamma_{m}$), grey data points to those in the
fluid/hexatic phase. We postpone the discussion of the data points at
$\Gamma < \Gamma_{m}$ and first concentrate on the crystalline regime
where a renormalization of Lam$\acute{e}$ coefficients really makes
sense.  The thick dashed lines in Fig.~(\ref{fig2}) represent the
$T=0$ calculation which holds down to $\Gamma$ values close to $\Gamma
= 75$. The thick solid line shows the theoretical curve for
$\mu_R(\Gamma)$ and $\mu_R(\Gamma) + 2 \lambda_R(\Gamma)$, which we
computed following the renormalization procedure outlined in the NH
paper (Eq. (2.42),(2.43) and (2.45) in \cite{nelson79} with $K(\Gamma)
= 1.258\,\Gamma$ at $T=0$ and $\Gamma_m =60$ as input to set up the
boundary conditions.) Theory and experiment agree well, considering
that no fit parameter has been used. For $\Gamma>\Gamma_m$, all our
results are converged, meaning that the computed moduli do not depend
on the length of the trajectory.  This is demonstrated by means of the
$\Gamma=100$ measurement for which the moduli were computed taking
only the first third of all configurations (open square symbol in
Fig.~(\ref{fig2})).

\begin{figure}
\includegraphics[width=0.48 \textwidth]{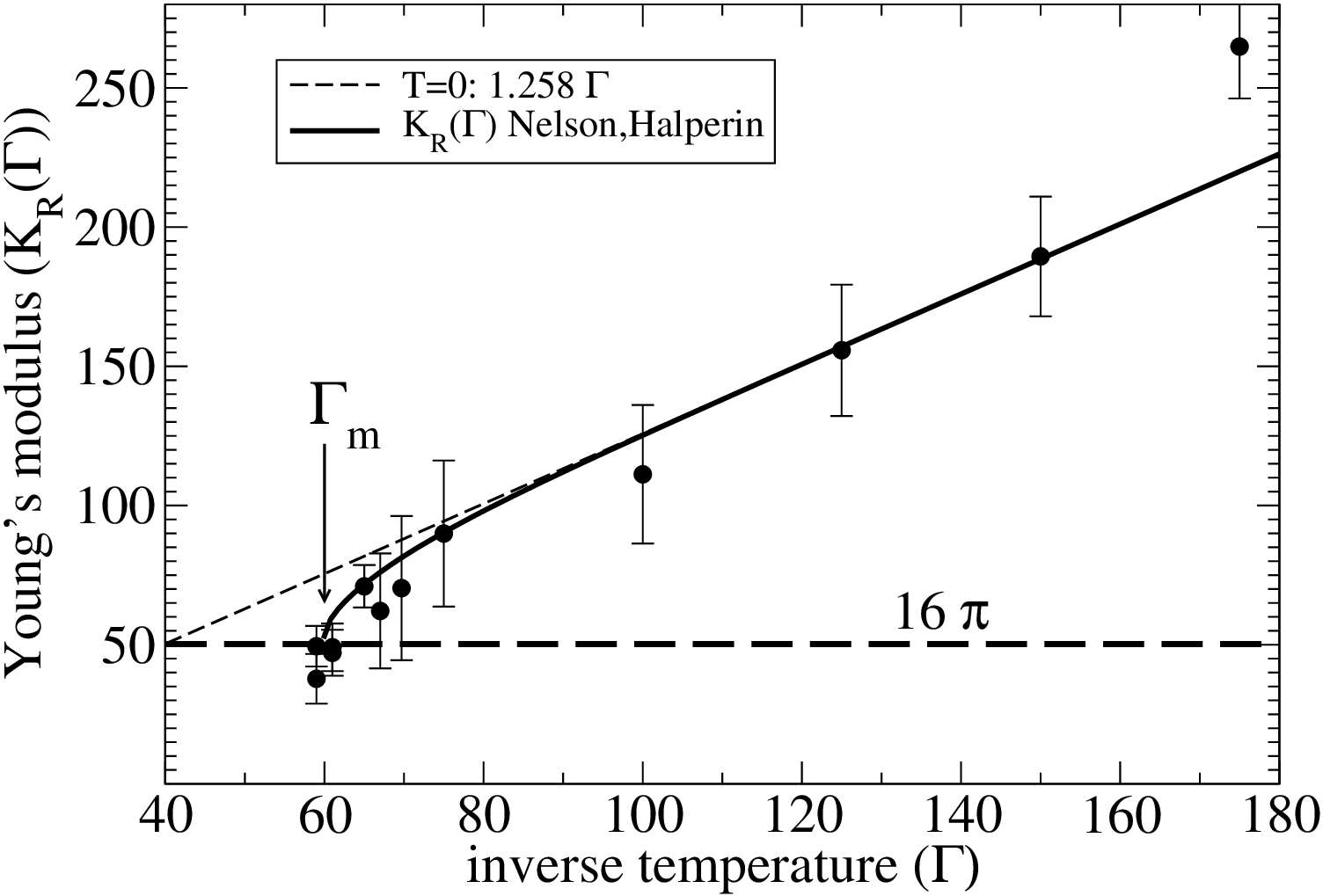}
\caption{\label{fig3}
  Young's modulus, eq.~(\ref{eq:0}), as a function of the inverse
  temperature, evaluated with the experimentally determined
  Lam$\acute{e}$ coefficients of Fig.~(\ref{fig2}) (symbols). The
  solid curve is $K_{R}(\Gamma)$ renormalized according to
  \cite{nelson79}, while the dashed curve is based on the T=0
  prediction.}
\end{figure}

Fig.~(\ref{fig3}) now checks eq.~(\ref{eq:0}), with $K_R(\Gamma)$
evaluated using the elastic moduli from Fig.~(\ref{fig2}). Using the
theoretical values from the $T=0$ calculation, we obtain $K(\Gamma) =
1.258\,\Gamma$, shown in Fig.~(\ref{fig3}) as dashed line. The thick
solid line shows the theoretical curve for $K_R(\Gamma)$ which we
computed with Lam$\acute{e}$ coefficients that were renormalized
following the NH procedure explained above. The main result of this
work is that the experimental data points closely follow the
theoretical curve and indeed, they cross $16 \pi$ at $\Gamma = \Gamma_m$
in excellent agreement with the predictions of NH. The length of the
remaining error bars correlate with the total measurement time, and
the $q$-range chosen in the extrapolation step.

The data points for $\Gamma < \Gamma_m$ in Fig.~(\ref{fig1}) and
(\ref{fig2}) should be treated with utmost caution. For the reasons
given above, it is not clear to us whether or not
eqs.~(\ref{eq:5},\ref{eq:6}) is at all meaningful outside the
crystalline phase. But even if it were, we should be aware that the
results will always depend on the measurement time.  This is
demonstrated for a system at $\Gamma=49$ for which the moduli were
calculated taking again just a fraction of 1/3 of all configurations
(open symbols in Fig.~(\ref{fig2})). There is almost an order of
magnitude difference to the data points based on all configurations,
thus indicating the dependence of the moduli on the length of the
analyzed trajectories. Physically, one could interpret this in terms of
a frequency-dependent shear modulus which for non-zero $\omega \sim
1/t_{exp}$ is known to exist even in fluids.

To conclude, we have measured particle trajectories of a
two-dimensional colloidal model system and computed elastic dispersion
curves which at low $\vec{q}$ give access to the elastic constants. We
thus measured $\mu_R$, $\lambda_R$ and Young's modulus $K_R$ as a
function of the inverse temperature $\Gamma$. All three quantities
compare well with corresponding predictions of the KTNHY
theory. Young's modulus, in particular, tends to $16 \pi$ when the
crystal melts, as predicted in \cite{nelson79}.

We acknowledge stimulating discussions with Matthias Fuchs and David
Nelson as well as financial support from the Deutsche
Forschungsgemeinschaft (European Graduate College 'Soft Condensed
Matter' and Schwerpunktprogramm Ferrofluide, SPP 1104)


\begin{thebibliography}{10}
\bibitem{Kosterlitz73} J. Kosterlitz and D. Thouless,
J. Phys. C: Solid State Phys., {\bf 6}, 1181 (1973).
\bibitem{mermin} N.D. Mermin, Phys. Rev. {\bf 176}, 250, (1968); R..E. Peierls,
  Helv. Phys. {\bf 7}, 81, (1923); L.D. Landau, Phys. Z. Sowjet, {\bf 11}, 26, (1937).
\bibitem{Halperin78} B. Halperin and D. Nelson, Phys. Rev. Lett.,
{\bf 41}, 121 (1978).
\bibitem{nelson79} D.R. Nelson and B.I. Halperin,
Phys. Rev. B, {\bf 19}, 2457 (1979).
\bibitem{Young79} A. Young, Phys. Rev. B,{\bf 19}, 1855 (1979).
\bibitem{Strandburg88}
K.~J. Strandburg, Rev. Mod. Phys., {\bf 60}, 161 (1988).
\bibitem{Glaser93}
M. Glaser and N. Clark, Adv. Chem. Phys., {\bf 83}, 543 (1993).
\bibitem{Grimes79}
C. Grimes and G. Adams, Phys. Rev. Lett., {\bf 42}, 795 (1979).
\bibitem{Morf79} R. Morf, Phys.Rev.Lett., {\bf 43}, 931 (1979).
\bibitem{Pieranski80} P. Pieranski,Phys. Rev. Lett., {\bf 45}, 569 (1980).
\bibitem{Murray87}C.A. Murray and D.H. Van Winkle, Phys. Rev. Lett.,
{\bf 58}, 1200 (1987).
\bibitem{Kusner94}
R. Kusner, J. Mann, J. Kerins, and A. Dahm,
Phys. Rev. Lett. {\bf 73},3113 (1994).
\bibitem{Marcus97}
A. Marcus and S. Rice, Phys. Rev. E, {\bf 55}, 637 (1997).
\bibitem{zahn99} K. Zahn, R. Lenke, and
  G. Maret, Phys. Rev. Lett. {\bf 82}, 2721 (1999)
\bibitem{Chen95}
K. Chen, T. Kaplan, and M. Mostoller
Phys. Rev. Lett., {\bf 74}, 4019 (1995).
\bibitem{Bagchi96}
K. Bagchi and H. Andersen, Phys. Rev. Lett., {\bf 76}, 255 (1996).
\bibitem{Jaster98}
A. Jaster, Europhys. Lett., {\bf 42}, 277 (1998).
\bibitem{Fisher82}
D. Fisher, Phys. Rev. B, {\bf 26}, 5009 (1982).
\bibitem{Morales94}
J. Morales, Phys. Rev. E {\bf 49}, 5127 (1994).
\bibitem{Sengupta00}
S. Sengupta, P. Nielaba, M. Rao and K. Binder, Phys. Rev. E, {\bf 61},
  1072 (2000);S. Sengupta, P. Nielaba and K. Binder Phys. Rev. E, {\bf
  61}, 6294 (2000).
\bibitem{zahn97} K. Zahn, J.M.Mendez-Alcaraz and G. Maret,
Phys. Rev. Lett {\bf 79}, 175 (1997); K. Zahn, G. Maret, C. Ru\ss{},
and H.H. von Gr\"unberg, Phys. Rev. Lett. {\bf 91}, 115502 (2003);
K. Zahn, A. Wille, G. Maret, S. Sengupta, and P. Nielaba,
Phys. Rev. Lett. {\bf 90}, 155506 (2003).
\bibitem{zahn00} K. Zahn and G. Maret, Phys. Rev. Lett. {\bf 85}, 3656
  (2000).
\bibitem{keim04} P. Keim, G. Maret, U. Herz, and H.H. von Gr\"unberg,
Phys.Rev.Lett., {\bf 92},215504-1 (2004).
\bibitem{foot37HN} see footnote 37 of \cite{nelson79}.
\bibitem{wille}   A.Wille, Ph.D. thesis, University of Konstanz,
  Konstanz, Germany,   (2001).
\end{thebibliography}
\end{document}